\documentclass[conference]{IEEEtran}
\usepackage{blindtext, graphicx,amsmath}

\newcommand{\remove}[1]{}

\newcommand{\denseSpace}{\renewcommand{\baselinestretch}{0.80}\normalsize}
\newcommand{\normalSpace}{\renewcommand{\baselinestretch}{1.0}\normalsize}
\usepackage[lined, ruled, linesnumbered, noresetcount]{algorithm2e}

\newcommand{\lref}[1]{line~\NlSty{\ref{#1}}}

\newcommand{\llref}[2]{lines~\NlSty{\ref{#1}}--\NlSty{\ref{#2}}}

\SetKwBlock{Loop}{loop}{end loop}
\SetKwData{Int}{\textbf{int}}
\SetKwData{Bool}{\textbf{boolean}}
\SetKw{Array}{array}
\SetKw{Of}{of}
\SetKw{Const}{const}
\SetKw{Define}{define}
\SetKw{Init}{init}
\SetKw{Goto}{goto}
\SetKw{Await}{await}
\SetKw{True}{true}
\SetKw{False}{false}
\SetKw{Cont}{continue loop}
\SetKw{Local}{local}
\SetKw{Shared}{shared}

\usepackage{standalone}
\usepackage[dvipsnames]{xcolor}
\usepackage{graphicx}
\usepackage{pgfplots}

\usepackage{tikz}
\usepackage{pgfplotstable}
\usepackage{caption}
\usepackage{multicol}

\usetikzlibrary{pgfplots.groupplots}

\makeatletter

\def\addlegendimage{\pgfplots@addlegendimage}
\makeatother

\begin{document}
\title{Node-Centric Detection of Overlapping Communities in Social Networks}
\author{\IEEEauthorblockN{Yehonatan Cohen, Danny Hendler, Amir Rubin}
\IEEEauthorblockA{Computer Science Department, Ben-Gurion University of the Negev, Be'er-Sheva, Israel\\
\{yehonatc,hendlerd,amirrub\}@cs.bgu.ac.il}

\remove{
\and
\IEEEauthorblockN{Danny Hendler}
\IEEEauthorblockA{Computer Science Department, Ben-Gurion University of the Negev, Be'er-Sheva, Israel\\
Email: hendlerd@cs.bgu.ac.il}
\and
\IEEEauthorblockN{Amir Rubin}
\IEEEauthorblockA{Computer Science Department\\
Ben-Gurion University of the Negev\\
Be'er-Sheva, Israel\\
Email: amirrub@cs.bgu.ac.il}}
}

\maketitle

\begin{abstract}
We present NECTAR, a community detection algorithm that generalizes Louvain method's local search heuristic for overlapping community structures. NECTAR chooses dynamically which objective function to optimize based on the network on which it is invoked. Our experimental evaluation on both synthetic benchmark graphs and real-world networks, based on ground-truth communities, shows that NECTAR provides excellent results as compared with state of the art community detection algorithms.
\end{abstract}

\begin{IEEEkeywords}
Community detection, overlapping communities, objective function, modularity, Louvain method.
\end{IEEEkeywords}

\IEEEpeerreviewmaketitle

\section{Introduction}
Social networks tend to exhibit community structure \cite{fortunato2010community}, that is, they may be partitioned to sets of nodes called \emph{communities} (a.k.a. \emph{clusters}), each of which relatively densely-interconnected, with relatively few connections between different communities. Revealing the community structure underlying complex networks in general, and social networks in particular, is a key problem with many applications (see e.g. \cite{krogan2006global,flake2002self}) that is the focus of intense research. Numerous community detection algorithms were proposed
(see e.g. \cite{le2013fast,lancichinetti2009detecting,prat2014high,esquivel2011compression,
blondel2008fast,newman2004finding,xie2012towards,gregory2010finding,adamcsek2006cfinder,
lancichinetti2011finding,gregory2011fuzzy}). While research focus was initially on detecting \textit{disjoint communities}, in recent years there is growing interest in the detection of \textit{overlapping communities}, where a node may belong to several communities. Indeed, social networks often possess overlapping community structure, since people are  members of multiple communities.

Many community detection algorithms are guided by an\textit{ objective function} that provides a quality measure of the clusterings they examine in the course of their execution. Since exhaustive-search optimization of these functions is generally intractable (see e.g. \cite{DBLP:conf/wg/BrandesDGGHNW07,DBLP:conf/sofsem/SimaS06}), existing methods settle for an approximation of the optimum and employ heuristic search strategies.

A key example is Blondel et al.'s algorithm \cite{blondel2008fast}, also known by the name "Louvain method" (LM). The algorithm is fast and relatively simple to understand and use and has been successfully applied for detecting  communities in numerous networks. It aims to maximize the modularity objective function \cite{newman2004finding}. Underlying the algorithm is a greedy local search heuristic that iterates over all nodes, assigning each node to the community it fits most (as quantified by modularity) and seeking a local optimum. Unfortunately, for reasons we elaborate on later, the applicability of LM is limited to disjoint community detection. In this work we show that LM's simple local search heuristic can be generalized in a natural manner to obtain a highly effective detector for overlapping communities.

\subsection{Our Contributions}

We present NECTAR, a Node-centric ovErlapping Community deTection AlgoRithm. NECTAR generalizes the node-centric local search heuristic of the widely-used Louvain method \cite{blondel2008fast} so that it can be applied also for networks possessing overlapping community structure.

Several algorithmic issues have to be dealt with in order to allow the LM heuristic to support multiple community-memberships per node. First, rather than adding a node $v$ to the \textit{single} community maximizing an objective function, $v$ may have to be added to several such communities. However, since the ``correct'' number of communities to which $v$ should belong is not a-priori known to the algorithm, it must be chosen dynamically.

A second issue that arises from multiple community-memberships is that different communities with large overlaps may emerge during the algorithm's execution and must be merged. We provide a detailed description of the new algorithm and how it resolves these issues in Section \ref{sec:NECTAR}.

Modularity (used by LM) assumes disjoint communities. Which objective functions should be used for overlapping community detection? Yang and Leskovec \cite{yang2015defining} evaluated several objective functions and showed that which is most appropriate depends on the network at hand. They observe that objective functions that are based on triadic closure provide the best results when there is significant overlap between communities. Weighted Community Clustering (WCC) \cite{prat2012shaping} is such an objective function but is defined only for disjoint community structures.

We present Weighted Overlapping Community Clustering (WOCC), a generalization of WCC that may be applied for overlapping community detection (described in Section \ref{sec:NECTAR}). Another objective function that fits the overlapping setting is $Q^E$ - an extension of modularity for overlapping communities \cite{chen2014extension} that, as indicated by the results of our experiments, is more adequate for graphs with relatively small inter-community overlap.

A unique feature of NECTAR is that it chooses dynamically whether to use WOCC or $Q^E$, depending on the structure of the graph at hand. This allows NECTAR to provide good results on graphs with both high and low community overlaps. To the best of our knowledge, NECTAR is the first community-detection algorithm that selects dynamically which objective function to use based on the graph on which it is invoked.

Local search heuristics guided by an objective function may be categorized as either \emph{node-centric} or \emph{community-centric}. Node-centric heuristics iterate over nodes. For each node, communities are considered and it is added to those of them that are ``best'' in terms of the objective function. Community-centric heuristics do the opposite. They iterate over communities. For each community, nodes are considered and the ``best'' nodes are added to it. In order to investigate which of these approaches is superior in the context of social networks, we implemented both a node-centric and a community-centric versions of NECTAR and compared the two implementations using both the WOCC and the $Q^E$ metrics. Our results show that the node-centric approach was significantly superior for both metrics.

We conducted extensive competitive analysis of NECTAR (using a node-centric approach) and six other state-of-the-art overlapping community detection algorithms.  Our evaluation was done using both synthetic graphs and real-world networks with ground-truth communities. We evaluated the clusterings output by the algorithms using several commonly-used metrics. NECTAR outperformed all other algorithms in terms of average detection quality and was best or second-best for almost all networks. 

The rest of this paper is organized as follows. We survey key related work in Section \ref{sec:related}. We provide detailed description of NECTAR in Section \ref{sec:NECTAR}. We report on our experimental evaluation in Section \ref{sec:evaluation}.
We compare the node-centric and community-centric approaches in Section \ref{node-community-centric}. We conclude with a discussion in Section \ref{sec:discussion}.

\section{Related Work}
\label{sec:related}
In this section, we briefly describe a few key notions directly related to our work.
Blondel et al.'s algorithm \cite{blondel2008fast}, a.k.a. the Louvain method, is a widely-used disjoint community detection algorithm. It is based on a simple node-centric search heuristic that seeks to maximize \emph{modularity} \cite{newman2004finding} - a global objective function that estimates the quality of a graph partition. For a given partitioning $\cal{C}$, modularity is defined as: $Q({\cal{C}}) = \sum_{C_i \in {\cal{C}}} (e_{i,i}-{a_i}^2)$, where $e_{i,j}$ is the fraction of edges in the network that connect vertices
in community $C_i$ to those in community $C_j$, and $a_i = \sum_{j : C_j \in {\cal{C}}} (e_{i,j})$.

Chen et al. extended the definition of modularity to the overlapping setting \cite{chen2014extension}.
For a collection of sets of nodes $\cal{C}$, their \emph{extended modularity} definition, denoted $Q^E(\cal{C})$, is given by:
\begin{equation}
\label{eq:emod}
Q^E({\cal{C}}) = \frac{1}{2|E|}\sum\limits_{C\in {\cal{C}}}\sum\limits_{i,j\in C} \bigg[ A_{ij}-\frac{k_i k_j}{2|E|} \bigg] \frac{1}{O_i O_j},
\end{equation}
where $A$ is the adjacency matrix, $k_i$ is the degree of node $i$, and $O_i$ is the number of communities $i$ is a member of. If $\cal{C}$ is a partition of network nodes, $Q^E$ reduces to (regular) modularity.

Yang and Leskovec \cite{yang2015defining} conducted a comparative analysis of $13$ objective functions in order to determine which of them captures better the community structure of a network. They show that which function is best depends on the network at hand. They also observe that objective functions that are based on \emph{triadic closure} provide the best results when there is significant overlap between communities.

Weighted Community Clustering (WCC) \cite{prat2012shaping} is such an objective function. It is based on the observation that triangle structures are much more likely to exist within communities than across them. This observation is leveraged for quantifying the quality of graph partitions (that is, non-overlapping communities). It is formally defined as follows. For a set of nodes $S$ and a node $v$, let $t(v,S)$ denote the number of triangles that $v$ closes with nodes of $S$. Also, let $vt(v,S)$ denote the number of nodes of $S$ that form at least one triangle with $v$. $WCC(v,S)$, quantifying the extent by which $v$ should be a member of $S$, is defined as:
\begin{equation}
\label{eq:contribxS}
WCC(v,S)=\left\{
\begin{array}{ll}
\frac{t(v,S)}{t(v,V)} \cdot \frac{vt(v,V)}{\vert S \setminus {v} \vert + vt(v,V \setminus S)} & \text{if }t(v,V) > 0\\
0 & \text{otherwise},
\end{array} \right.
\end{equation}
where $V$ is the set of graph nodes. The cohesion level of a community $S$ is defined as:
\begin{equation}
\label{eq:communityCohesion}
WCC(S)=\frac{1}{\vert S \vert} \sum_{v \in S} WCC(v,S).
\end{equation}

Finally, the quality of a partition ${\cal{C}}=\{C_1,\ldots,C_n\}$ is defined as the following weighted average:
\begin{equation}
\label{eq:partitionQuality}
WCC({\cal{C}})=\frac{1}{\vert V \vert} \sum_{i=1}^{n} \vert C_i \vert \cdot WCC(C_i).
\end{equation}

In Section \ref{sec:NECTAR}, we present Weighted Overlapping Community Clustering (WOCC) - a generalization of WCC that can serve as an objective function for overlapping community detection.

\section{NECTAR: a Detailed Description}
\label{sec:NECTAR}

The high-level pseudo-code of NECTAR is given by Algorithm \ref{fig:Alg1}. The input to the \texttt{NECTAR} procedure (see \lref{NECTAR:start}) is a graph $G=<V,E>$ and an algorithm parameter $\beta \geq 1$ that is used to determined the number of communities to which a node should belong in a dynamic manner (as we  describe below).

\texttt{NECTAR} proceeds in iterations (\llref{NECTAR:iterLoopStart}{NECTAR:iterLoopEnd}), which we call \emph{external iterations}. In each external iteration, the algorithm performs \emph{internal iterations}, in which it iterates over all nodes $v \in V$ (in some random order), attempting to determine the set of communities to which node $v$ belongs such that the objective function is maximized. NECTAR selects dynamically whether to use WOCC or $Q^E$, depending on the rate of closed triangles in the graph on which it is invoked. If the average number of closed triangles per node in $G$ is above the \emph{trRate} threshold, then WOCC is more likely to yield good performance and it is used, otherwise the extended modularity objective function is used instead (\llref{ifHighTriRate}{useEmod}). We use $\text{\emph{trRate}}=5$, as this provides a good separation between communities with high overlap (on which WOCC is superior) and low overlap (on which extended modularity is superior). We elaborate on the two objective functions and their implications on the algorithm in Section \ref{section:obj-functions}.

Each internal iteration (comprising \llref{NECTAR:nodesDo}{NECTAR:nodesDoEnd}) proceeds as follows. First, \texttt{NECTAR} computes the set $C_v$ of communities to which node $v$ currently belongs (\lref{NECTAR:computeCv}). Then, $v$ is removed from all these communities (\lref{NECTAR:removeFromCv}). Next, the set $S_v$ of $v$'s neighboring communities (that is, the communities of $\cal{C}$ that contain one or more neighbors of $v$) is computed in \lref{NECTAR:Sv}. Then, the gain in the objective function value that would result from adding $v$ to each neighboring community (relative to the current set of communities $\cal{C}$) is computed in \lref{NECTAR:computeDeltas}. Node $v$ is then added to the community maximizing the gain in objective function and to any community for which the gain is at least a fraction of $1/\beta$ of that maximum (\llref{NECTAR:computeC'v}{NECTAR:addToC'v}).\footnote{If no gain is positive, $v$ remains as a singleton.}
Thus, the number of communities to which a node belongs may change dynamically throughout the computation, as does the set of communities $\cal{C}$.

If the internal iteration did not change the set of communities to which $v$ belongs, then $v$ is a \emph{stable node} of the current external iteration and the number of stable nodes (which is initialized to $0$ in \lref{NECTAR:initStable}) is incremented (\llref{NECTAR:noChange}{NECTAR:incrementStable}). After all nodes have been considered, the possibly-new set of communities is checked in order to prevent the emergence of different communities that are too similar to one another. This is accomplished by the \texttt{merge} procedure (whose code is not shown), called in \lref{NECTAR:merge}. It receives as its single parameter a value $\alpha$ and merges any two communities whose relative overlap is $\alpha$ or more. More precisely, each pair of communities $C_1, C_2 \in \cal{C}$ is merged if $\vert C_1 \cap C_2 \vert / min\{|C_1|,|C_2|\} \geq \alpha$ holds. We use $\alpha=0.8$, as this is the value that gave the best results (\lref{Var:alpha}). If the number of communities was reduced by \texttt{merge}, the counter of stable nodes is reset to $0$ (\llref{NECTAR:Ifmerged}{NECTAR:resetS}).

The computation proceeds until either the last external iteration does not cause any changes (hence the number of stable nodes equals $\vert V \vert$) or until the maximum number of iterations is reached (\lref{NECTAR:RepeatCondition}), whichever occurs first. We have set the maximum number of iterations to $20$ (\lref{Var:maxIter}) in order to strike a good balance between detection quality and runtime. In practice, the algorithm converges within a fewer number of iterations in the vast majority of cases. For example, in our experiments on synthetic graphs with $5000$ nodes, NECTAR converges after at most $20$ iterations in $99.5$\% of the executions.

LM is a hierarchical clustering algorithm that has a second phase, in which a new network is constructed whose nodes are the communities discovered in the first phase. The weights of the edges between these nodes are given by the total weights of links between the corresponding communities.  The algorithm is then re-invoked on the new network. We implemented a hierarchical version of NECTAR. However, in all our experiments, the best results were obtained in the first hierarchy level. Consequently, in the descriptions and evaluation results that follow, we refer to the non-hierarchical version of NECTAR (Algorithm \ref{fig:Alg1}) unless stated otherwise.

\denseSpace
\NoCaptionOfAlgo
\begin{algorithm}[t]
\DontPrintSemicolon

\caption{\textbf{Figure 1: NECTAR algorithm pseudo-code. \label{fig:Alg1}}}

\Const $\text{maxIter}$ $\gets$ 20 \tcc*{max iterations} \nllabel{Var:maxIter}
\Const $\alpha$ $\gets$ 0.8 \tcc*{merge threshold} \nllabel{Var:alpha}
\Const $trRate$ $\gets$ 5 \tcc*{WOCC threshold} \nllabel{Var:trRate}
\BlankLine

\textbf{Procedure} \texttt{NECTAR}(\emph{G=$<$V,E$>$}, $\beta$)\{ \nllabel{NECTAR:start}\;
\uIf{$triangles(G)/\vert V \vert \geq  trRate$ \nllabel{ifHighTriRate}}
    {use WOCC \tcc*{use WOCC obj. function}}
\Else{use $Q^E$ \tcc*{use $Q^E$ obj. function} \nllabel{useEmod}}
Initialize communities \nllabel{NECTAR:init}\\

$i \gets 0$ \tcc*{number of extern. iterations}

\Repeat{$(s = \vert V \vert) \lor (i = maxIter)$ \nllabel{NECTAR:RepeatCondition}}
    {\nllabel{NECTAR:iterLoopStart}
    $s \gets 0$ \tcc*{number of stable nodes} \nllabel{NECTAR:initStable}
    \ForAll{$v \in V$ \nllabel{NECTAR:nodesDo}}
        {
        $C_v$ $\gets$ communities to which $v$ belongs \nllabel{NECTAR:computeCv}\\
        Remove $v$ from all the communities of $C_v$ \nllabel{NECTAR:removeFromCv}\;
        $S_v \gets \{C \in {\cal{C}} \big{\vert} \exists u: u \in C \land (v,u) \in E \}$ \nllabel{NECTAR:Sv}\;
        $D \gets \{\Delta(v,C) \vert C \in S_v \}$ \nllabel{NECTAR:computeDeltas}\;
        $C'_v \gets \{C \in S_v \vert \Delta(v,C) \cdot \beta \geq max(D) \}$ \nllabel{NECTAR:computeC'v}\;
        Add $v$ to all the communities of $C'_v$ \nllabel{NECTAR:addToC'v}\;
        \uIf{$C'_v = C_v$ \nllabel{NECTAR:noChange}}
            {
            $s$++ \nllabel{NECTAR:incrementStable}
            }
        } \nllabel{NECTAR:nodesDoEnd}

            \texttt{merge}($\alpha$) \tcc*{merge communities} \nllabel{NECTAR:merge}
            \uIf{\emph{merge reduced number of communities} \nllabel{NECTAR:Ifmerged}}
                {
                $s \gets$0 \nllabel{NECTAR:resetS}
                }
        $i$++\;

    } \nllabel{NECTAR:iterLoopEnd}

\end{algorithm}
\normalSpace

\subsection{Objective Functions}
\label{section:obj-functions}
As mentioned previously, we implemented the extended modularity function \cite{chen2014extension}, denoted $Q^E(C)$, and WOCC - a generalization of the WCC function \cite{prat2012shaping}. NECTAR decides dynamically which of these functions to use based on the rate of triangles in the graph on which it is invoked.

The implementation of the $\Delta$ function, used in \llref{NECTAR:computeDeltas}{NECTAR:computeC'v}, as well as that of the initialization (\lref{NECTAR:init}), is different depending on the objective function used. We now describe these implementation details.

\subsubsection*{Extended Modularity}
The extended modularity function is given by Equation \ref{eq:emod} (see Section \ref{sec:related}). However, for the purposes of computing the $\Delta$ function, a clustering is quantified as follows:
\begin{equation}
\label{eq:AWDelta}
\sum\limits_{i\in c} \bigg[ A_{iv}-\frac{k_i k_v}{2|E|} \bigg] \frac{1}{O_i}.
\end{equation}

The expression above is derived from Equation \ref{eq:emod} as follows. The $\frac{1}{2|E|}$ factor is removed; as it is the same for all $C \in S_v$, removing it does not change $C'_v$. The first summation in Equation \ref{eq:emod} is also removed, since we consider the gain in $Q^E$ that results from adding $v$ to a specific community $C$. Finally, since $v$ is removed from all communities before the gains are computed, $O_v$ equals $1$.
The initialization in \lref{NECTAR:init} is done by simply setting $C_v=\{v\}$ for all $v \in V$.

\subsubsection*{Weighted Overlapping Community Clustering}

The WCC objective function is given by Equations \ref{eq:contribxS}-\ref{eq:partitionQuality} (see Section \ref{sec:related}).
We use a generalization of it that we call Weighted Overlapping Community Detection (WOCC). Unlike WCC, WOCC supports weighted edges and multiple community-memberships per node. Since edge weights were required only by the hierarchical version of NECTAR, we do not describe here the changes in WCC for supporting them here, except for noting that when all edge weights are $1$, $WCC(v,S)=WOCC(v,S)$ holds. 

The computation of $WOCC(\cal{C})$, which is the value of the WOCC objective function for partition $\cal{C}$, is done as in Equation \ref{eq:partitionQuality}, except that the left-hand factor is $1 / (\sum_{C \in {\cal{C}}} \vert C \vert)$ instead of $\frac{1}{\vert V \vert}$. This is required in order to account for multiple community-memberships per node.

Initialization for WOCC is done as in \cite{prat2014high}. We consider the nodes in decreasing order of their clustering coefficient. For each node $v$, if not placed in a community already, we construct a new community containing $v$ and all its neighbors not in a community already.

\section{Experimental Evaluation}
\label{sec:evaluation}
Xie et al. \cite{xie2013overlapping} conducted a comparative study of state-of-the-art overlapping community detection algorithms. We compare the performance of NECTAR with that of the following $5$ of the key performers out of the $14$ algorithms they evaluated.

The \emph{Greedy Clique Expansion} (GCE) algorithm \cite{lee2010detecting}, due to Hurly et al., is an agglomerative algorithm. It uses maximal cliques (whose size $k$ is given as an algorithm parameter) as its seeds and expands them in a greedy manner. Similar communities are merged.
The \emph{Cfinder} algorithm \cite{adamcsek2006cfinder} uses $k$-size cliques (where $k$ is an algorithm parameter) as its seeds and then  merges all communities sharing at least $k-1$ nodes. This is an implementation of the well-known Clique Percolation Method (CPM) \cite{palla2005uncovering}. The \emph{Order Statistics Local Optimization Method} (OSLOM) agglomerative algorithm \cite{lancichinetti2011finding} is due to Lancichinetti et al. It identifies communities by maximizing a local fitness function on their nodes. OSLOM uses statistical tools to estimate cluster significance.  It receives the value of the significance threshold as a parameter.

The \emph{Community Overlap PRopagation Algorithm} (COPRA) by Gregory employs the label propagation technique for community detection \cite{raghavan2007near}. Nodes contain labels that propagate along edges so that nodes can reach agreement on their community membership. Each node may belong to up to $v$ communities, where $v$ is an algorithm parameter.
The \emph{Speaker-Listener Label Propagation Algorithm} (SLPA) \cite{xie2012towards}, due to Xie et al., uses the label propagation technique as well. Similarly to COPRA, SLPA accounts for overlap by allowing each node to possess multiple community-labels but different features are used to control community membership. An algorithm threshold parameter $r$ is used in the final step of label selection.

In addition to the top performers of \cite{xie2013overlapping}, we also evaluate the \emph{Fuzzy-Infomap} algorithm \cite{gregory2011fuzzy}, due to Gregory. The algorithm extends Infomap \cite{rosvall2008maps} to deal with overlapping communities. It considers \emph{fuzzy memberships}, in which nodes may belong to different communities to different extents, unlike \emph{crisp memberships}, where each node fully belongs to each community of which it is a member.

We used the following algorithm parameters. When NECTAR invokes WOCC, we use $12$ different values of $\beta$ in the range $[1.1,20]$. When it invokes $Q^E$, we use $13$ different values of $\beta$ in the range $[1.01,1.4]$. For GCE, we used $k \in \{3,\dots,8\}$. We used Cfinder's default setting, in which it starts with seeds of size varying between $3$ and the size of the maximum graph clique. For OSLOM, we perform $10$ executions, in each of which different nodes are randomly selected to be the seeds of communities. For COPRA, we used $v \in \{1,2,\dots,10\}$, performing $10$ executions per every value of $v$ and choosing the execution yielding maximum modularity. For SLPA, we used the default setting, which performs $11$ executions with varying values of parameter $r$ in the range $[0.01,0.5]$. For Fuzzy Infomap, we have set the flag indicating overlapping communities.

We conducted competitive analysis using both synthetic networks and real-world networks with ground-truth. We evaluated results using several commonly-used metrics. Our evaluation shows that NECTAR outperformed all other algorithms in terms of average detection quality and provided best or second-best results for almost all networks. We now describe the evaluation criteria we use. This is followed by details on our experiments and their results.

\subsubsection*{Evaluation Criteria}
The evaluation criteria we use assume the existence of the ground-truth cover for the analysed graph. This is indeed the case for the synthetic graphs and real-world networks on which we conduct our experiments. We quantify the quality of the cover computed by the algorithms by employing the following widely-used measures.

\begin{enumerate}

\item \emph{Normalized Mutual Information} (NMI) \cite{lancichinetti2009detecting} is based on the notion of normalized mutual information and uses entropy to quantify the extent by which we may learn about one cover given the other and is defined as follows.
$$NMI({\cal{C}}_1,{\cal{C}}_2) = 1- \frac{1}{2}(H({\cal{C}}_1|{\cal{C}}_2) + H({\cal{C}}_2|{\cal{C}}_1)),$$
where $H({\cal{C}}_1|{\cal{C}}_2)$ is the conditional entropy of cover ${\cal{C}}_1$ w.r.t. cover ${\cal{C}}_2$. As mentioned in \cite{mcdaid2011normalized}, in cases where one cover contains many more communities than the other, NMI is not a good representation of a cover's quality. We will return to this issue when estimating the quality of a cover given the ground-truth for real-world networks.

\item \emph{Omega-index} \cite{collins1988omega} is based on the fraction of pairs that occur together in the same number of communities in both covers, with respect to the expected value of this fraction in the null model. Unlike NMI, this measure refers to the nodes and the relationships between them, giving us a different view on a cover's quality w.r.t. ground-truth. We use the following version of Omega-index, used in \cite{xie2013overlapping}.\\
$\omega({\cal{C}}_1,{\cal{C}}_2) = \frac{\omega_ u({\cal{C}}_1,{\cal{C}}_2)- \omega_ e({\cal{C}}_1,{\cal{C}}_2)}{1-\omega_ e({\cal{C}}_1,{\cal{C}}_2)}$\\
$\omega_ u({\cal{C}}_1,{\cal{C}}_2) = \frac{1}{\binom{n}{2}} \sum\limits_{j=0}^{min(\vert {\cal{C}}_1 \vert,\vert {\cal{C}}_2 \vert)} |t_j({\cal{C}}_1) \cap t_j({\cal{C}}_2)|$\\
$\omega_ e({\cal{C}}_1,{\cal{C}}_2) = \frac{1}{\binom{n}{2}^2} \sum\limits_{j=0}^{min(\vert {\cal{C}}_1 \vert,\vert {\cal{C}}_2 \vert)} |t_j({\cal{C}}_1)|\cdot |t_j({\cal{C}}_2)|$\\

$t_j({\cal{C}}) = \{(x,y): |\{C \in {\cal{C}} :x,y \in C\}| = j\}$\\

\item \emph{Average F1 score} ($\bar{F_1}({\cal{C}}_1,{\cal{C}}_2)$), as presented in \cite{yang2012community}: For each community in the ground-truth and in the evaluated cover, we find the community in the other cover with the highest F1 score (in terms of node community-membership), where $F1(C_1,C_2)$ is the harmonic mean of precision and recall between node-sets $C_1$, $C_2$. We then compute the average score for ground-truth communities and the average score for the evaluated cover and compute their average.
$precision(C_1,C_2)=\frac{|C_1 \cap C_2 |}{|C_1|}$\\
$recall(C_1,C_2)=\frac{|C_1 \cap C_2 |}{|C_2|}$\\
$H(a,b) = \frac{2 \cdot a \cdot b}{a+b}$\\
$F_1(C_1,C_2) = H(precision(C_1,C_2),recall(C_1,C_2))$\\
$F_1(C_1,{\cal{C}}) = max\{F_1(C_1,C_i) : C_i \in {\cal{C}}\}$\\
And $\bar{F_1}({\cal{C}}_1,{\cal{C}}_2)$ is set to be:\\
$\frac{1}{2|{\cal{C}}_1|} \sum\limits_{C_i \in {\cal{C}}_1} F_1(C_i,{\cal{C}}_2) + \frac{1}{2|{\cal{C}}_2|} \sum\limits_{C_i \in {\cal{C}}_2} F_1(C_i,{\cal{C}}_1)$\\

\end{enumerate}

\subsubsection*{Synthetic Networks}
Lancichinetti et al. \cite{lancichinetti2008benchmark} introduced a set of benchmark graphs (henceforth the LFR benchmark) that provide heterogeneity in terms of node degree and community-size distributions, as well as control of the degree of overlap.
We mostly use the same LFR parameter values used by \cite{xie2013overlapping}, as follows. The number of nodes, $n$, is set to $5000$. The average node degree, $k$, is set to either $10$ or $40$, and the number of overlapping nodes, $O_n$, is set to either $10\%$ or $50\%$ of the total number of nodes, respectively. The number of communities an overlapping node belongs to, $O_m$, is set to values in the range $\{2, \ldots, 8\}$. The exponent for degrees distribution, $\tau_1$, is set to $2$ and the exponent for community size distribution, $\tau_2$, is set to $1$. The maximum degree is set to $50$ and the mixing parameter (the expected fraction of links through which a node connects to nodes outside its communities), $\mu$, is set to $0.3$.

As for community sizes, the options are either big communities, whose size varies between $20-100$, or small communities, whose size varies between $10-50$. As done in \cite{xie2013overlapping}, we generate $10$ instances for each combination of parameters. We take the average of the results for each algorithm and each metric over these $10$ instances. For each algorithm, we present the results for the algorithm parameter value that maximizes this average.

\begin{figure}
	\begin{center}
		\includegraphics[width=\columnwidth]{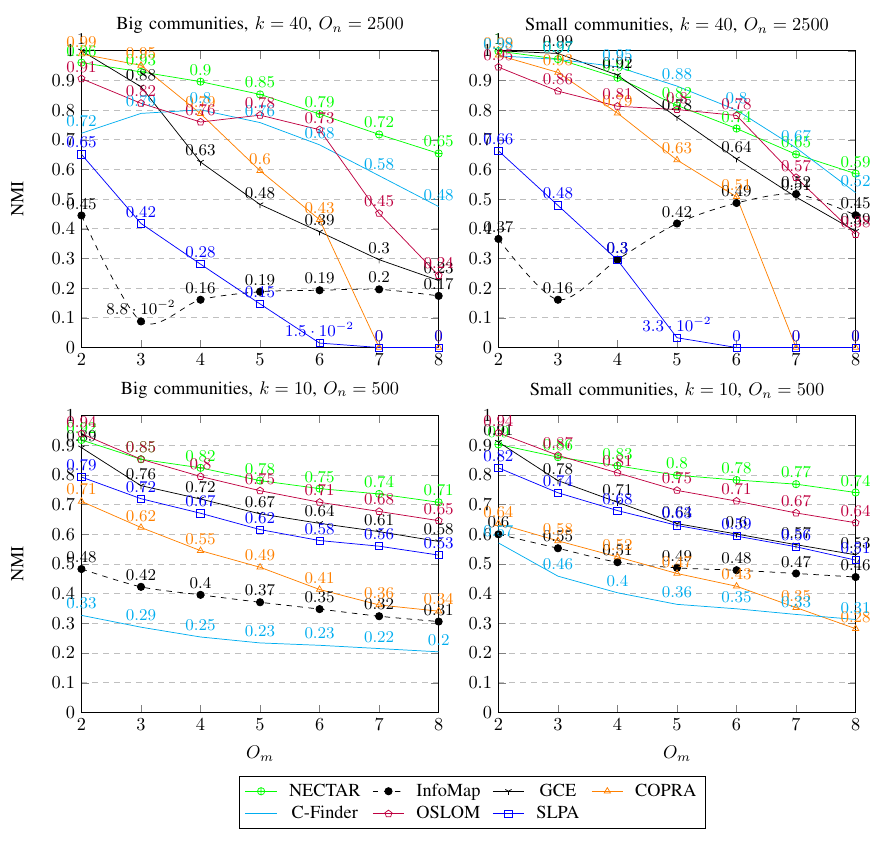}
	\end{center}
	\caption{NMI.}
	\label{fig:NMI}
\end{figure}

Figure \ref{fig:NMI} presents the average performance of the algorithms in terms of NMI as a function of $O_m$ (the number of communities to which each of the $O_n$ overlapping nodes belongs), for $k \in \{10,40\}$ and $O_n \in \{2500,5000\}$. The Omega-index and average F1 score results follow the same trends and are therefore omitted.

With only a few exceptions, it can be seen that the performance of the algorithms decreases as $O_m$ increases. This can be attributed to the fact that the size of the solution space increases with $O_m$.

We focus first on the results on graphs with a higher number of overlapping nodes ($O_m = 2500$) and high average degrees ($k=40$). The rate of triangles in these graphs is high (approx. $30$) and so NECTAR employs WOCC.
NECTAR is the clear leader for big communities. It achieves the best results for almost all values of $O_m$ and its relative performance improves as $O_m$ increases, confirming that the combination of NECTAR's search strategy and the WOCC objective function is suitable for graphs with significant overlap. COPRA takes the lead for $O_m \in \{2,3\}$ but then declines sharply. Cfinder improves its relative performance as $O_m$ increases and is the second performer for $O_m \in \{4,7,8\}$.
For small communities, Cfinder has the lead with NECTAR being second best and OSLOM third for most values of $O_m$, and NECTAR taking the lead for $O_m=8$.

We now describe the results on graphs with lower numbers of overlapping nodes ($O_m = 500$) and low average degree ($k=10$). The rate of triangles in these graphs is low (approx. $3.5$) and so NECTAR employs extended modularity. NECTAR provides the best performance for both small and large communities for almost all values of $O_m$. The relative performance of Cfinder deteriorates as compared with its performance on high-overlap graphs. It is not optimized for sparser graphs, since its search for communities is based on locating cliques. OSLOM is second best on these graphs, having the upper hand for $O_m=1$ and providing second-best performance for $O_m>1$. These results highlight the advantage of NECTAR's capability of selecting the objective function it uses dynamically according to the properties of the graph at hand.

Summarizing the results of the tests we conducted on $96$ different synthetic graph types, NECTAR is ranked first among the $7$ algorithms, with average rank of $1.58$, leading in $33$ out of $96$ of the tests, followed by OSLOM, with average rank of $2.79$. When looking only at the graphs with high overlapping rates (in which nodes with multiple communities are at least at $5$ communities, e.g. $O_m>4$), NECTAR's average rank improves to $1.31$, followed again by OSLOM ranked $2.78$ on average.

\subsubsection*{Real-World Networks}
We conducted our competitive analysis on two real-world networks - Amazon's product co-purchasing network and the DBLP scientific collaboration network. We downloaded both datasets from Stanford's Large Network Dataset Collection \cite{snapnets}. The Amazon graph consists of $334,863$ nodes and $925,872$ edges. Nodes represent products and edges are between commonly co-purchased products. The set of products from the same category is viewed as a ground-truth community.

\noindent The DBLP graph consists of $317,080$ nodes and $1,049,866$ edges.
Nodes correspond to authors and edges connect authors that have co-authored a paper. Publication venues (specifically, conferences) are used for defining ground-truth communities. Thus, the set of authors that have published in the same conference is viewed as a ground-truth community.

In \cite{yang2015defining}, Yang and Leskovec rate the quality of ground-truth communities of Amazon and DBLP (as well as those of additional networks) using six scoring functions, such as modularity, conductance, and cut ratio. They rank ground-truth communities based on the average of their ranks over the six corresponding scores and maintain the $5,000$ top ground-truth communities per each network. These are the ground-truth communities provided as part of the datasets of \cite{snapnets}.

As mentioned previously, the evaluation criteria we use return meaningful results when applied to a pair of

\begin{figure}
	\begin{center}
		\includegraphics[width=\columnwidth]{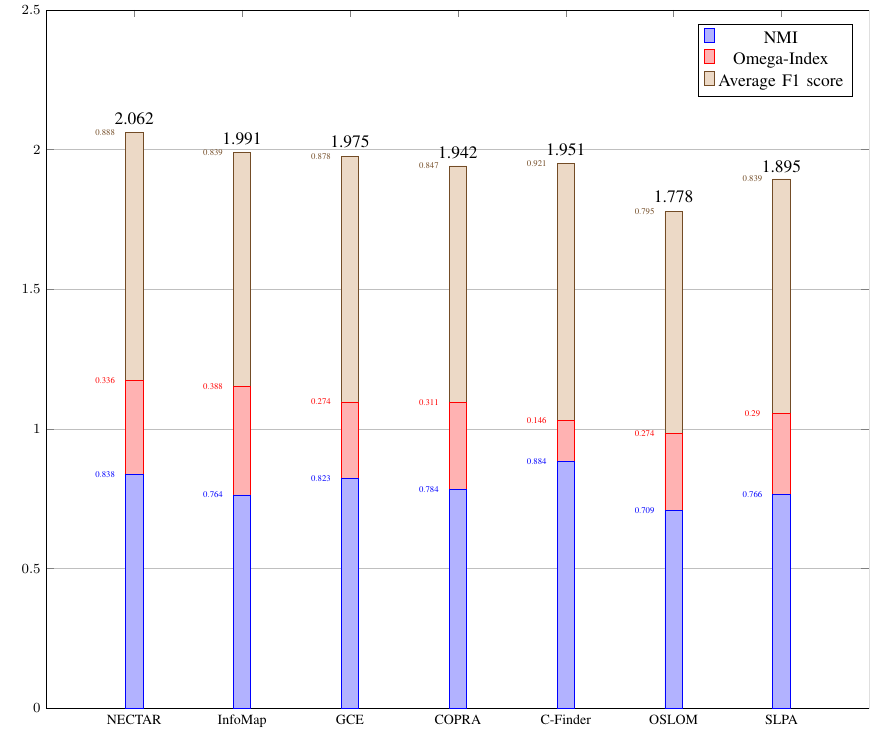}
	\end{center}
	\caption{Amazon competitive analysis.}
	\label{fig:Amazon}
\end{figure}

\noindent covers of more-or-less the same size. We guarantee this as follows. Let $\cal{G}$ be the set of ground-truth communities and let $\cal{C}$ be the cover produced by the algorithm. For each ground-truth community, we select a community in $\cal{C}$ that is most similar to it. More formally, for each ground-truth community $G \in \cal{G}$, we select a single community $C=argmax\{F1(G,C') : C' \in \cal{C}\}$. This results in a subset ${\cal{D}} \subset \cal{C}$ of size at most $\vert \cal{G}\vert$. (The size of $\cal{D}$ may be smaller than that of $\cal{G}$ since duplicates are eliminated.) We now apply the NMI, Omega-index and average F1 score criteria to quantify the quality of $\cal{D}$.

Figure \ref{fig:Amazon} presents the results of our competitive analysis on Amazon. Each bar corresponds to an algorithm and shows its scores on the $3$ evaluation criteria. The rate of triangles in the Amazon graph is low, and so NECTAR employs extended modularity. NECTAR provides the best performance with an overall score of $2.062$, approximately $3.5\%$ more than InfoMap, which is second best, and approximately $4.4\%$ more than GCE, which is the third performer. Zooming in to specific quality criteria, NECTAR is second-best in terms of NMI and average F1 score, lagging only slightly behind Cfinder in both cases. In terms of Omega-index, NECTAR is second-best as well, lagging behind InfoMap, and Cfinder is the last performer.

\begin{figure}
	\begin{center}
		\includegraphics[width=\columnwidth]{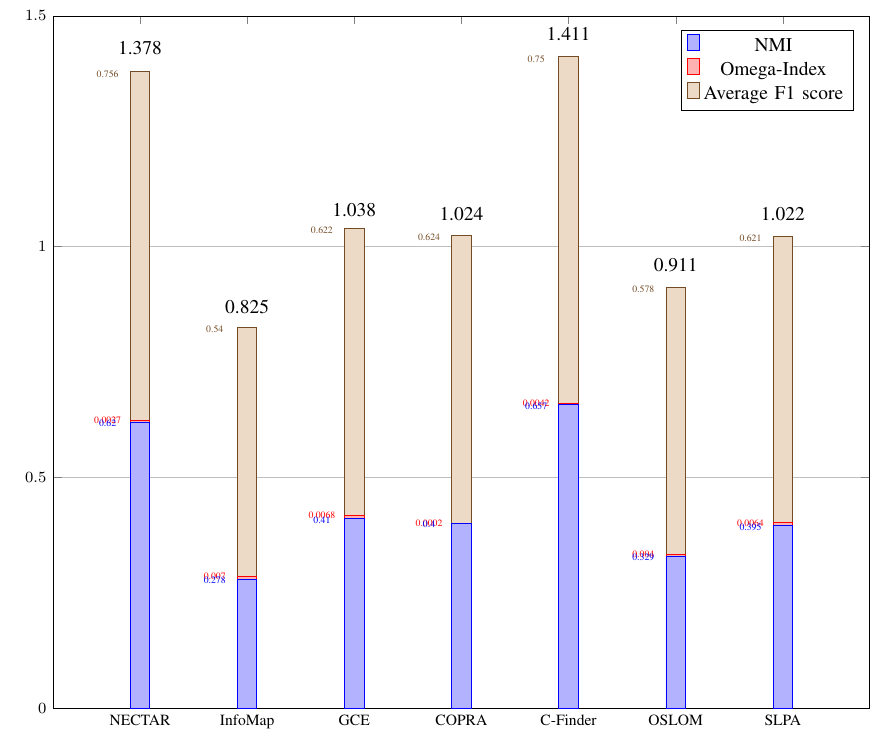}
	\end{center}
	\caption{DBLP competitive analysis.}
	\label{fig:DBLP}
\end{figure}

Figure \ref{fig:DBLP} presents the results on DBLP. The rate of triangles in DBLP is high, and so NECTAR employs WOCC. All algorithms fair poorly in terms of their Omega-index. In terms of overall score, Cfinder is the best performer, enjoying a small margin of approximately $2.5\%$ w.r.t. NECTAR, which is second-best.
GCE is the third performer, with a score lower than NECTAR's by approximately $25\%$. In terms of NMI, Cfinder is first with a score of $0.657$. NECTAR is second best, lagging by approximately $5.5\%$, and the third performer by a wide margin is GCE. In terms of average F1 score, NECTAR comes out first, but Cfinder's score is only approximately $1\%$ smaller. COPRA obtains the third score, nearly $17\%$ less than NECTAR's.

\section{Node-Centric vs. Community-Centric Search}
\label{node-community-centric}

While node-centric heuristics iterate over nodes, trying to find the best communities from each node's perspective, community-centric heuristics iterate over communities, trying to find the best nodes to add to each community. Out of the algorithms we experimented with, NECTAR is node-centric, while GCE and OSLOM are community-centric. Although the rest of the algorithms do not fall strictly into any of these categories, community memberships are nevertheless decided from either a node or a community viewpoint. For example, in SLPA and COPRA, which are label-propagation algorithms, community-membership is decided from a node's perspective.

NECTAR's greedy local search heuristic is node-centric. Since it is not a-priori clear which approach is superior, we decided to implement and evaluate a community-centric version of NECTAR as well. The community-centric version was implemented as follows. Instead of iterating over nodes (as done by NECTAR in the loop of \llref{NECTAR:nodesDo}{NECTAR:nodesDoEnd}), we iterate over communities. For each community $C$,
we add to $C$ those neighboring nodes that contribute the most in terms of the objective function (using $\beta$ as the threshold parameter as we did in Algorithm \ref{fig:Alg1}). After some nodes are added to $C$, the bond of other nodes to $C$ may weaken, so we also perform a ``clean-up'' routine for removing such nodes.

When optimizing $Q^E$, unlike in the node-centric approach, we need to take into account $O_v$ - the number of communities that a neighboring node $v$ of $C$ belongs to, and so the expression in Equation \ref{eq:AWDelta} has to be divided by $O_v$ (in addition to being divided by $O_i$).

We compared the performance of (the node-centric) NECTAR with that of the community-centric variant using the same set of LFR graph types describe above. For each graph, NECTAR first selects the appropriate objective function (see \llref{ifHighTriRate}{useEmod} of Algorithm \ref{fig:Alg1}) and then the node-centric/community-centric code optimizes the selected function.

\begin{figure}
	\begin{center}
		\includegraphics[width=\columnwidth]{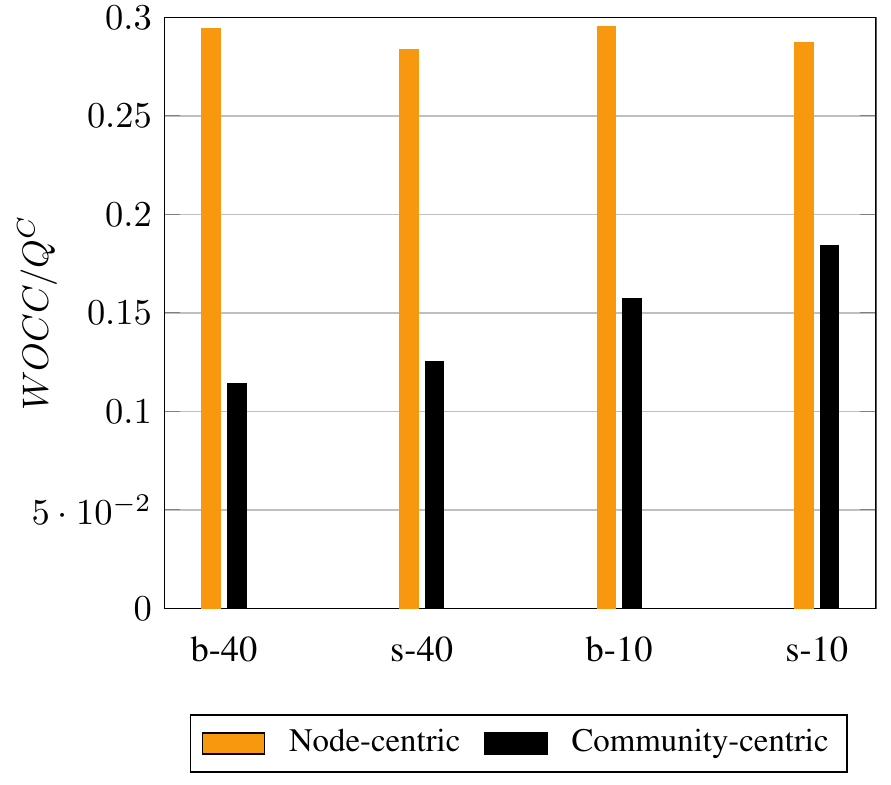}
	\end{center}
	\caption{Node-centric vs. community-centric.}
	\label{fig:node-comm}
\end{figure}

Figure \ref{fig:node-comm} presents the results. The (node-centric) NECTAR obtains values of the objective function that are, on average, more than twice those of the community-centric variant. Specifically, NECTAR exceeds the community-based variant by 72\% when extended modularity is optimized (graphs with average degree of $10$) and by 164\% when WOCC is optimized (graphs with average degree of $40$).

Obviously, these results do not necessarily mean that the node-centric approach is always superior to the community-centric approach, as this may depend on the graph at hand, the objective function optimized, and the implementation details of the search heuristic. Nevertheless, our intuition is that node-centric search may yield better results on networks in which the dynamics of community emergence are node-centric. In other words, node-centric search seems more natural for networks in which the agents themselves (represented by nodes) decide, either explicitly or implicitly, with whom to interact.

\section{Discussion}
\label{sec:discussion}
We introduced NECTAR, a novel overlapping community detection algorithm that generalizes the local search heuristic of the Louvain method so that it can be applied to networks possessing overlapping community structure.

A unique feature of NECTAR is that it selects dynamically which objective function to optimize, depending on the structure of the graph at hand. Yang and Leskovec \cite{yang2015defining} observe that objective functions that are based on triadic closure provide the best results when there is significant overlap between communities. Weighted Community Clustering (WCC) \cite{prat2012shaping} is such an objective function but is defined only for disjoint community structures. We define WOCC - a generalization of WCC that may be applied to overlapping communities. NECTAR uses WOCC when it is invoked on graphs that possess a high rate of closed triangles, whereas, for graphs with a low rate of closed triangles, it optimizes extended modularity \cite{shen2009detect} instead.

We conducted extensive experimental evaluation of NECTAR and six other state-of-the-art overlapping community detection algorithms.  Our evaluation was done using both synthetic graphs and real-world networks with ground-truth. We evaluated the clusterings output by the algorithms using several commonly-used metrics.
NECTAR outperformed all other algorithms in terms of average detection quality and was best or second-best for almost all networks.

Analysis of our empirical results shows that extended modularity yields better results on networks with low average node degrees and low community overlap, whereas WOCC yields better results on networks with higher degrees and overlap. The fact that NECTAR is able to provide excellent results on both types of networks highlights the importance of objective function dynamic selection, as well as the general applicability of Louvain's search heuristic.

NECTAR employs a node-centric heuristic that iterates over nodes, trying to find the best communities from each node's perspective. Some community detection algorithms take a different, community-centric approach, by iterating over communities, trying to find the best nodes to add to each community. Since it is not a-priori clear which approach is superior, we implemented a community-centric version of NECTAR and evaluated it using the LFR benchmark. Our evaluation shows that the node-centric approach was significantly superior on all LFR graph types.

This work opens up several interesting directions for future work. First, we plan to further investigate mechanisms for dynamic selection of the objective function. NECTAR chooses between WOCC and extended modularity, depending on the rate of closed triangles. In general, however, it might be possible to improve detection accuracy further, by selecting from a wider variety of objective functions, based on additional structural graph properties. Moreover, search heuristics that target some weighted average of several objective functions, instead of selecting just one of them, seem a promising approach.

Another direction for future work is to gain a more complete understanding, from both empirical and theoretical perspectives, of performance tradeoffs between node-centric and community-centric search heuristics. Questions that come to mind in this context are under what circumstances is one superior to the other, and whether they can be combined in a useful manner.

Finally, although most community detection algorithms require one or more user-provided parameters, eliminating such parameters, or at least reducing their number, simplifies their usage. We will seek ways of making NECTAR parameter-free.

\bibliographystyle{IEEEtran}
\bibliography{Main_File}

\end{document}